\documentclass{article}
\usepackage{epsfig}
\usepackage{graphicx}
\usepackage{amsmath}
\usepackage{amssymb}
\usepackage{fancybox}
\usepackage{geometry}

\begin{document}

\title{Construction of Lumps with nontrivial interaction}
\author{P. G. Est{\'e}vez\\
Facultad de Ciencias. Universidad de Salamanca.\\
Salamanca, 37008, Salamanca. Spain.\\
e-mail: pilar@usal.ess}
\date{}
 \maketitle

\abstract{We develop a method based upon the Singular Manifold Method that yields  an iterative and analytic procedure to construct solutions for a Bogoyavlenskii-Kadomtsev-Petviashvili equation. This method allows us to construct a rich collection of lump solutions with a nontrivial evolution behavior}
\section{Introduction}
In recent years, it has been proven in several papers \cite {Estevez:Fokas1983}, \cite{Estevez:Villarroel1999}, \cite{Estevez:Ablowitz2000} that the KPI equation contains a whole manifold of smooth rationally decaying ``lump" configurations associated with higher-order pole meromorphic eigenfunctions. These configurations have an interesting  dynamics and the lumps may scatter in a nontrivial way. Furthermore, algorithmic methods, based upon the Painlev\'e property, have been developed in order to construct lump-type solutions for different equations such as a $2+1$ NLS  (Nonlinear Schr\"odinger equation) \cite{Estevez:Estevez2007}, \cite{Estevez:Villarroel2009} equation and the KPI (Kadomtsev-Petviashvili equation)  and GDLW (Generalized Dispersive Wave Equation) equations \cite{Estevez:Estevez2008}.

The present contribution is related to the construction of lump solutions for the $2+1$ dimensional equation \cite {Estevez:Yu1998}
\begin{equation}\label{Estevez:equation1}
(4u_{xt}+u_{xxxy}+8u_xu_{xy}+4u_{xx}u_y)_x+\sigma u_{yyy}=0,\quad \sigma=\pm 1
\end{equation}
which represents a modification of the Calogero-Bogoyavlenskii-Schiff (CBS) equation \cite {Estevez:Calogero1975}, \cite{Estevez:Bogoyavlenskii1990}, \cite{Estevez:Schiff1992}:
\[ 4u_{xt}+u_{xxxy}+8u_xu_{xy}+4u_{xx}u_y=0 . \]

 Equation (\ref{Estevez:equation1}) has  often been called the Bogoyavlenskii-Kadomtsev-Petviashvili (KP-B) equation\cite{Estevez:Estevez2000}.

 As in the case of the KP equation, there are two versions of (\ref{Estevez:equation1}), depending upon the sign of $\sigma$. Here we restrict ourselves  to the minus sign. Therefore:
 \begin{equation}\label{Estevez:equation2}
(4u_{xt}+u_{xxxy}+8u_xu_{xy}+4u_{xx}u_y)_x- u_{yyy}=0
\end{equation}
or
\begin{gather}
\nonumber 4u_{xt}+u_{xxxy}+8u_xu_{xy}+4u_{xx}u_y=\omega_{yy}
,\\
u_y=\omega_x.\label{Estevez:equation3}
\end{gather}

 We  refer to  (\ref{Estevez:equation3}) as KP-BI in what follows.

 In Section 2 we  summarize the results that the singular method provides for KP-BI. These results are not essentially new because they were obtained by the author in \cite{Estevez:Estevez2000} for the KP-BII version of the equation. Section 3 is devoted to the construction of rational solitons.

 \section{The Singular Manifold Method for KP-BI}
 It has been proven  that  (\ref{Estevez:equation3}) has the Painlev\'e property \cite{Estevez:Yu1998}. Therefore, the singular manifold method can be applied to  it. In this section we  adapt previous results obtained in \cite{Estevez:Estevez2000} for KP-BII  to KP-BI . This is why we are only present the main results with no detailed explanation since this has been shown in our earlier paper.
 \subsection{The singular Manifold Method}
 This method \cite{Estevez:Weiss1983} requires the truncation of the Painlev\'e series for the fields $u$ and $\omega$ of (\ref{Estevez:equation3})in the following form:
 \begin{gather}
\nonumber u^{[1]}=u^{[0]}+\frac{\phi^{[0]}_x}{\phi^{[0]}} ,\\
\omega^{[1]}=\omega^{[0]}+\frac{\phi^{[0]}_y}{\phi^{[0]}},\label{Estevez:equation4}
\end{gather}
where $\phi^{[0]}(x,y,t)$ is the singular manifold and $u^{[i]},\, \omega^{[i]}\, (i=0,1)$ are solutions of (\ref{Estevez:equation3}). This means that  (\ref{Estevez:equation4}) can be considered as an auto-B\"acklund transformation. The substitution of (\ref{Estevez:equation4}) in (\ref{Estevez:equation3}) yields a polynomial in negative powers of $\phi^{[0]}$ that can be   handled  with MAPLE. The result (see \cite{Estevez:Estevez2000})
is that we can express the seed solutions $u^{[0]},\, \omega^{[0]}$ in terms of the singular manifold as follows:

 \begin{gather}
\nonumber u^{[0]}_x=\frac{1}{4}\left(-v_x-\frac{v^2}{2}-z_y+\frac{z_x}{2}\right),\\
u^{[0]}_y
=\omega^{[0]}_x=\frac{1}{4}\left(-r-2v_y+2z_x z_y\right),\label{Estevez:equation5}
\end{gather}
where $v, r$ and $z$ are related to the singular manifold $\phi$ through the following definitions
 \begin{equation}
v=\frac{\phi^{[0]}_{xx}}{\phi^{[0]}_x},\quad
 r=\frac{\phi^{[0]}_t}{\phi^{[0]}_x}, \label{Estevez:equation6}\quad
 z_x=\frac{\phi^{[0]}_y}{\phi^{[0]}_x}.
\end{equation}
Furthermore, the singular manifold $\phi^{[0]}$ satisfies the equation
 \begin{equation}
s_y+r_x-z_{yy}-z_xz_{xy}-2z_yz_{xx}=0\label{Estevez:equation7}
\end{equation}
where $s=v_x-\frac{v^2}{2}$ is the Schwartzian derivative.
\subsection {Lax pair}
Equations  (\ref{Estevez:equation5}) can be linearized through the following definition of $\psi^{[0]}(x,y,t)$, $\chi^{[0]}(x,y,t)$  functions.
 \begin{gather}
\nonumber v=\frac{\psi^{[0]}_x}{\psi^{[0]}}+\frac{\chi^{[0]}_x}{\chi^{[0]}},\\
 z_x=i\left(\frac{\psi^{[0]}_x}{\psi^{[0]}}-\frac{\chi^{[0]}_x}{\chi^{[0]}}\right).\label{Estevez:equation8}
\end{gather}

When one combines (\ref{Estevez:equation5}), (\ref{Estevez:equation6}) and (\ref{Estevez:equation7}), the following Lax pair arises:
 \begin{gather}
\nonumber \psi^{[0]}_{xx}=-i\psi^{[0]}_y-2u^{[0]}_x\psi^{[0]}\\
\psi^{[0]}_t=2i\psi^{[0]}_{yy}-4u^{[0]}_y\psi^{[0]}_x+(2u^{[0]}_{xy}+2i\omega^{[0]}_y)\psi^{[0]}\label{Estevez:equation9}
\end{gather}
together with  its complex conjugate
 \begin{gather}
\nonumber \chi^{[0]}_{xx}=i\chi^{[0]}_y-2u^{[0]}_x\chi^{[0]}\\
\chi^{[0]}_t=-2i\chi^{[0]}_{yy}-4u^{[0]}_y\chi^{[0]}_x+(2u^{[0]}_{xy}-2i\omega^{[0]}_y)\chi^{[0]}.\label{Estevez:equation10}
\end{gather}
In terms of $\chi^{[0]}$ and $\psi^{[0]}$, the derivatives of $\phi^{[0]}$ are:
 \begin{gather}
\nonumber v=\frac{\phi^{[0]}_{xx}}{\phi^{[0]}_x}=\frac{\psi^{[0]}_x}{\psi^{[0]}}+\frac{\chi^{[0]}_x}{\chi^{[0]}}\Rightarrow \phi^{[0]}_x=\psi^{[0]}\chi^{[0]},\\
 r=\frac{\phi^{[0]}_t}{\phi^{[0]}_x}=-4u^{[0]}_y+2\frac{\psi^{[0]}_y}{\psi^{[0]}}\frac{\psi^{[0]}_x}{\psi^{[0]}}+2\frac{\psi^{[0]}_x}{\psi^{[0]}}\frac{\psi^{[0]}_y}{\psi^{[0]}}-2\frac{\psi^{[0]}_{xy}}{\psi^{[0]}}-2\frac{\chi^{[0]}_{xy}}{\chi^{[0]}}, \label{Estevez:equation11}\\
\nonumber z_x=\frac{\phi^{[0]}_y}{\phi^{[0]}_x}=i\left(\frac{\psi^{[0]}_x}{\psi^{[0]}}-\frac{\chi^{[0]}_x}{\chi^{[0]}}\right),
\end{gather}
which allows us to write  $d\phi^{[0]}$ as:
 \begin{gather}
\nonumber d\phi^{[0]}=\psi^{[0]}\chi^{[0]}\,dx+i\left(\chi^{[0]}\psi^{[0]}_x-\psi^{[0]}\chi^{[0]}_x\right)\,dy+\\ +\left(-4u^{[0]}_y\psi^{[0]}\chi^{[0]}+2\psi^{[0]}_y\chi^{[0]}_x+2\psi^{[0]}_x\chi^{[0]}_y-2\chi^{[0]}\psi^{[0]}_{xy}-2\psi^{[0]}\chi^{[0]}_{xy}\right)\,dt.\label{Estevez:equation12}
\end{gather}
It is easy to check that the condition of the exact derivative in (\ref{Estevez:equation12}) is satisfied by the Lax pairs (\ref{Estevez:equation9}) and (\ref{Estevez:equation10}).
\subsection{Darboux transformations}
Let $(\psi_1^{[0]},\chi_1^{[0]})$, $(\psi_2^{[0]},\chi_2^{[0]})$ be  two pairs of eigenfunctions of the Lax pair (\ref{Estevez:equation9})-(\ref{Estevez:equation10}) corresponding to the seed solution $u^{[0]},\, \omega^{[0]}$
 \begin{gather}
\nonumber \psi^{[0]}_{j,xx}=-i\psi^{[0]}_{j,y}-2u^{[0]}_x\psi_j^{[0]},\\
\psi_{j,t}^{[0]}=2i\psi^{[0]}_{j,yy}-4u^{[0]}_y\psi^{[0]}_{j,x}+(2u^{[0]}_{xy}+2i\omega^{[0]}_y)\psi^{[0]}_j,\label{Estevez:equation13}\\
\nonumber\\
\nonumber \chi^{[0]}_{j,xx}=i\chi^{[0]}_{j,y}-2u^{[0]}_x\chi^{[0]}_j,\\
\chi^{[0]}_{j,t}=-2i\chi^{[0]}_{j,yy}-4u^{[0]}_y\chi^{[0]}_{j,x}+(2u^{[0]}_{xy}-2i\omega^{[0]}_y)\chi^{[0]}_j,\label{Estevez:equation14}
\end{gather}
where $j=1,2$.
These Lax pairs can be considered as nonlinear equations between the fields and the eigenfunction \cite{Estevez:Estevez2000}. This means that the Painlev\'e expansion of the fields
 \begin{gather}
\nonumber u^{[1]}=u^{[0]}+\frac{\phi^{[0]}_{1,x}}{\phi^{[0]}_1} ,\\
\omega^{[1]}=\omega^{[0]}+\frac{\phi^{[0]}_{1,y}}{\phi^{[0]}_1},\label{Estevez:equation15}
\end{gather}
should be accompanied by an expansion of the eigenfunctions and the singular manifold itself. These expansions are
\begin{gather}
\nonumber \psi^{[1]}_2=\psi^{[0]}_2-\psi^{[0]}_1\,\frac{\Omega_{1,2} }{\phi^{[0]}_1},\\
\chi^{[1]}_2=\chi^{[0]}_2-\chi^{[0]}_1\,\frac{\Omega_{2,1} }{\phi^{[0]}_1},\label{Estevez:equation16}\\
\nonumber\phi^{[1]}_2=\phi^{[0]}_2-\frac{\Omega_{1,2} \Omega_{2,1} }{\phi^{[0]}_1}.
\end{gather}
Substitution of (\ref{Estevez:equation16}) in (\ref{Estevez:equation13}-\ref{Estevez:equation14}) yields
 \begin{gather}
\nonumber d\Omega_{i,j}=\psi^{[0]}_j\chi^{[0]}_i\,dx+i\left(\chi^{[0]}_i\psi^{[0]}_{j,x}-\psi^{[0]}_j\chi^{[0]}_{i,x}\right)\,dy+\\ +\left(-4u^{[0]}_y\psi^{[0]}_i\chi^{[0]}_j+2\psi^{[0]}_{j,y}\chi^{[0]}_{i,x}+2\psi^{[0]}_{j,x}\chi^{[0]}_{i,y}-2\chi^{[0]}_i\psi^{[0]}_{j,xy}-2\psi^{[0]}_j\chi^{[0]}_{i,xy}\right)\,dt.\label{Estevez:equation17}
\end{gather}
Direct comparison of (\ref{Estevez:equation12}) and (\ref{Estevez:equation17}) affords
$\phi^{[0]}_j=\Omega_{j,j}$.
Therefore,  knowledge of the two seed eigenfunctions $(\psi^{[0]}_j,\chi^{[0]}_j),\,j=1,2,$ allows us to compute the matrix elements $\Omega_{i,j}$, which yields the Darboux transformation (\ref{Estevez:equation15}-\ref{Estevez:equation16}).
\subsection{Iteration: $\tau$-functions}
According to the above results, $\phi^{[1]}_2$ is a singular manifold for the iterated fields $u^{[1]},\, \omega^{[1]}$. Therefore, the Painlev\'e expansion for these iterated fields can be written as
 \begin{gather}
\nonumber u^{[2]}=u^{[1]}+\frac{\phi^{[1]}_{2,x}}{\phi^{[1]}_2} ,\\
\omega^{[2]}=\omega^{[1]}+\frac{\phi^{[1]}_{2,y}}{\phi^{[1]}_2},\label{Estevez:equation18}
\end{gather}
which combined with (\ref{Estevez:equation15}) is:
 \begin{gather}
\nonumber u^{[2]}=u^{[0]}+\frac{(\tau_{1,2})_x}{\tau_{1,2}} ,\\
\omega^{[2]}=\omega^{[0]}+\frac{(\tau_{1,2})_y}{\tau_{1,2}}\label{Estevez:equation19},
\end{gather}
where $\tau_{1,2}=\phi^{[1]}_2\phi^{[0]}_1$,
which according to (\ref{Estevez:equation15}) allows us to write it as
\begin{equation}
\tau_{1,2}=\phi^{[0]}_2\phi^{[0]}_1-\Omega_{1,2}\Omega_{2,1}=\det(\Omega_{i,j}).\label{Estevez:equation20}\end{equation}

\section{Lumps}
The  iteration method described above can be started from the most trivial initial solution $u^{[0]}=\omega^{[0]}=0$. In this case, the lax pair, is:
 \begin{gather}
\nonumber \psi^{[0]}_{j,xx}=-i\psi^{[0]}_{j,y},\\
\nonumber\psi{[0]}_{j,t}=2i\psi^{[0]}_{j,yy},\\
\nonumber \chi^{[0]}_{j,xx}=i\chi^{[0]}_{j,y},\\
\chi^{[0]}_{j,t}=-2i\chi^{[0]}_{j,yy}.\label{Estevez:equation21}
\end{gather}
It is  trivial  to prove that equations (\ref{Estevez:equation21}) have the following solutions
\begin{gather}
\nonumber\psi^{[0]}_1=P_m(x,y,t;k)\,\exp\,\left[{Q_0(x,y,t;k)}\right],\\
\nonumber\chi^{[0]}_1=P_n(x,y,t;k)\,\exp\,\left[{-Q_0(x,y,t;k)}\right],\\
\psi^{[0]}_2=\left(\chi^{[0]}_1\right)^*,\label{Estevez:equation22}\\
\nonumber\chi^{[0]}_2=\left(\psi^{[0]}_1\right)^*,
\end{gather}
where $m,n$ are arbitrary integers  and $k$ an arbitrary complex constant.
\begin{equation}Q_0(x,y,t;k)=kx+ik^2y+2ik^4t\Rightarrow (Q_0(x,y,t;k))^*=k^*x-i(k^*)^2y-2i(k^*)^4t\label{Estevez:equation23}\end{equation}
and $P_j(x,y,t;k)$ is defined as:
\begin{equation}P_j(x,y,t;k)\,\exp\,\left[{Q_0(x,y,t;k)}\right]=\frac{\partial^j\left(P_{j-1}(x,y,t;k)\,\exp\,\left[{Q_0(x,y,t;k)}\right]\right)}{\partial k^j},\quad P_0=1.\label{Estevez:equation24}\end{equation}
These solutions are characterized by two integers, $n$ and $m$ that provide a rich collection of different solutions corresponding to the same wave number $k$. Thus in our opinion, all of them  should be considered as one-soliton solutions despite the different behaviors shown by the solutions corresponding to the different combinations of $n$ and $m$. We  now present some of these cases.
\subsection {Lump (0,0): $n=0,\quad m=0$}
\begin{figure}
\begin{center}
\epsfig{file=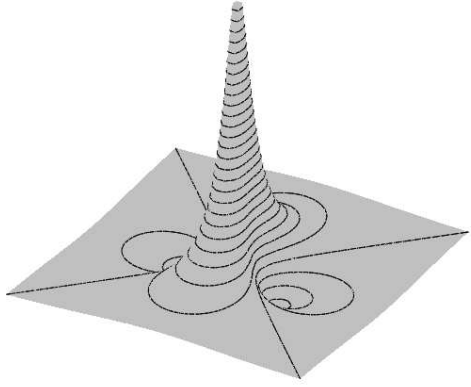,height=8 cm}
\end{center}
\caption{Lump of the $0,0$ type}
\end{figure}
The eigenfunctions (\ref{Estevez:equation22}) are:
\begin{gather}
\nonumber\psi^{[0]}_1=\exp\,\left[{Q_0(x,y,t;k)}\right],\\
\nonumber\chi^{[0]}_1=\exp\,\left[{-Q_0(x,y,t;k)}\right],\\
\psi^{[0]}_2=\exp\,\left[{-Q_0^*(x,y,t;k)}\right],\label{Estevez:equation25}\\
\nonumber\chi^{[0]}_2=\exp\,\left[{Q_0^*(x,y,t;k)}\right].
\end{gather}
The matrix elements (\ref{Estevez:equation17}) can be integrated as:
\begin{gather}
\nonumber \phi^{[0]}_1=\Omega_{1,1}= x+2iky-8ik^3t,\\
\nonumber \phi^{[0]}_2=\Omega_{2,2}=x-2ik^*y+8i\left(k^*\right)^3t,\\
\Omega_{1,2}= -\frac{1}{k+k^*}\exp\,\left[{-Q_0(x,y,t;k)}\right]\exp\,\left[{-Q_0^*(x,y,t;k)}\right],\label{Estevez:equation26}\\
\nonumber\Omega_{2,1}=\frac{1}{k+k^*}\exp\,\left[{Q_0(x,y,t;k)}\right]\exp\,\left[{Q_0^*(x,y,t;k)}\right].
\end{gather}
Therefore, the $\tau$-function (\ref{Estevez:equation20}) is the positive defined expression
\begin{equation}
\tau_{1,2}=X_1^2+Y_1^2+\frac{1}{4a_0^2},\label{Estevez:equation27}\end{equation}
where
\begin{gather}
\nonumber k=a_0+ib_0,\\
X_1=x-2b_0y+8b_0\left(3a_0^2-b_0^2\right)t,\label{Estevez:equation28}\\
\nonumber Y_1=2a_0\left(y+4(3b_0^2-a_0^2)t\right).
\end{gather}
The profile of this solution  is shown in Figure 1. It represents a lump (static in the variables $X_1, Y_1$) of height $8a_0^2$.

\subsection {Lump (1,0): $n=1,\quad m=0$}
The eigenfunctions (\ref{Estevez:equation22}) are:
\begin{gather}
\nonumber\psi^{[0]}_1=P_1(x,y,t;k)\,\exp\,\left[{Q_0(x,y,t;k)}\right],\\
\nonumber\chi^{[0]}_1=\exp\,\left[{-Q_0(x,y,t;k)}\right],\\
\psi^{[0]}_2=\exp\,\left[{-Q_0^*(x,y,t;k)}\right],\label{Estevez:equation29}\\
\nonumber\chi^{[0]}_2=\left\{P_1(x,y,t;k)\right\}^*\exp\,\left[{Q_0^*(x,y,t;k)}\right],
\end{gather}
where according to (\ref{Estevez:equation24}) we have:
\begin{equation}P_1(x,y,t;k)=x+2iky-8ik^3t.\label{Estevez:equation30}\end{equation}
In this case, the matrix elements (\ref{Estevez:equation17}) are:
\begin{gather}
\nonumber \phi^{[0]}_1=\frac{x^2}{2}+iy+2ixyk-2y^2k^2-12itk^2-8ixtk^3+16ytk^4-32t^2k^6=\\ \nonumber \quad = \frac{X_1^2-Y_1^2}{2}+i\frac{2a_0X_1Y_1+Y_1-16a_0^3t}{2a_0},\\
\nonumber \phi^{[0]}_2=\left(\phi^{[0]}_1\right)^*,\\
\Omega_{1,2}= -\frac{1}{2a_0}\,\exp\,\left[{-Q_0(x,y,t;k)}\right]\,\exp\,\left[{-Q_0^*(x,y,t;k)}\right],\label{Estevez:equation31}\\
\nonumber\Omega_{2,1}=\frac{1-2a_0X_1+2a_0^2\left(X_1^2+Y_1^2\right)}{4a_0^3}\,\exp\,\left[{Q_0(x,y,t;k)}\right]\,\exp\,\left[{Q_0^*(x,y,t;k)}\right].
\end{gather}
Therefore, the $\tau$-function (\ref{Estevez:equation20}) is the  positive defined expression
\begin{equation}
\tau_{1,2}=\left(\frac{X_1^2-Y_1^2}{2}\right)^2+\left(\frac{2a_0X_1Y_1+Y_1-16a_0^3t}{2a_0}\right)^2+\left(\frac{X_1-\frac{1}{2a_0}}{2a_0}\right)^2+\left(\frac{Y_1}{2a_0}\right)^2+\frac{1}{16a_0^4}.\label{Estevez:equation32}\end{equation}
The profile of this solution  is shown in Figure 2. If we wish to show its behavior when $t\rightarrow \pm\infty$, we need to look along the lines
\begin{gather}
\nonumber X_1=\hat X_1+c_1t ^{1/2}\\
 Y_1=\hat Y_1+c_2t ^{1/2}
\label{Estevez:equation33}
\end{gather}
such that  (\ref{Estevez:equation32}), when $t\rightarrow \pm\infty$, is different from $0$

$\bullet$ For $t<0$, the possibilities are $c_1=\pm 2a_0\sqrt{-2},\quad c_2=-c_1$ which yields two lumps approaching with opposite velocities along the lines
\begin{gather}
\nonumber X_1=\hat X_1\pm 2a_0(-2t)^{1/2}\\
 Y_1=\hat Y_1\pm 2a_0(-2t)^{1/2},
\label{Estevez:equation34}
\end{gather}
and the limit of $\tau_{1,2}$ along these lines is
\begin{equation}
\tau_{1,2}=\left(\hat X_1+\frac{1}{4a_0}\right)^2+\left(\hat Y_1-\frac{1}{4a_0}\right)^2+\frac{1}{4a_0^4}.\label{Estevez:equation35}\end{equation}

$\bullet$ For $t>0$, the possibilities are $c_1=\pm 2a_0\sqrt{2},\quad c_2=c_1$ which yields two lumps  with opposite velocities along the lines
\begin{gather}
\nonumber X_1=\hat X_1\pm 2a_0(2t)^{1/2}\\
 Y_1=\hat Y_1\pm 2a_0(2t)^{1/2},
\label{Estevez:equation36}
\end{gather}
and the limit of $\tau_{1,2}$ along these lines is
\begin{equation}
\tau_{1,2}=\left(\hat X_1+\frac{1}{4a_0}\right)^2+\left(\hat Y_1+\frac{1}{4a_0}\right)^2+\frac{1}{4a_0^4}.\label{Estevez:equation37}\end{equation}
\begin{figure}
\begin{center}
\epsfig{file=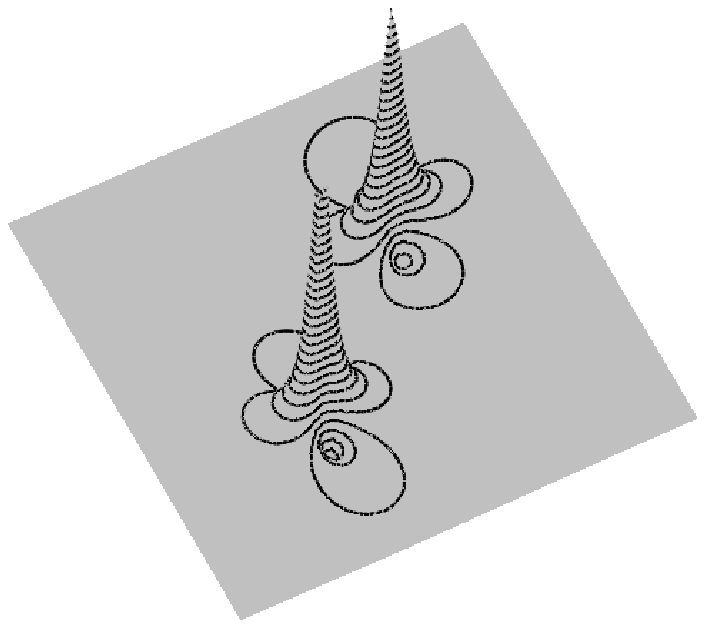,height=8cm}
\end{center}
\caption{Lump of the $1,0$ type}
\end{figure}
\subsection {Lump (1,1): $n=1,\quad m=1$}
The eigenfunctions (\ref{Estevez:equation22}) are:
\begin{gather}
\nonumber\psi^{[0]}_1=P_1(x,y,t;k)\,\exp\,\left[{Q_0(x,y,t;k)}\right],\\
\nonumber\chi^{[0]}_1=P_1(x,y,t;k)\exp\,\left[{-Q_0(x,y,t;k)}\right],\\
\psi^{[0]}_2=\left\{P_1(x,y,t;k)\right\}^*\exp\,\left[{-Q_0^*(x,y,t;k)}\right],\label{Estevez:equation38}\\
\nonumber\chi^{[0]}_2=\left\{P_1(x,y,t;k)\right\}^*\exp\,\left[{Q_0^*(x,y,t;k)}\right].
\end{gather}
This yields  the following  matrix elements according to (\ref{Estevez:equation17}):
\begin{gather}
\nonumber \phi^{[0]}_1= \frac{X_1\left(X_1^2-3Y_1^2\right)}{3}+i\left(X_1^2Y_1-\frac{Y_1^3}{3}+8a_0t\right),\\
\nonumber \phi^{[0]}_2=\left(\phi^{[0]}_1\right)^*,\\
\Omega_{1,2}= -\frac{1+2a_0X_1+2a_0^2\left(X_1^2+Y_1^2\right)}{4a_0^3}\,\exp\,\left[{-Q_0(x,y,t;k)}\right\}\,\exp\,\left[{-Q_0^*(x,y,t;k)}\right],\label{Estevez:equation39}\\
\nonumber\Omega_{2,1}=\frac{1-2a_0X_1+2a_0^2\left(X_1^2+Y_1^2\right)}{4a_0^3}\,\exp\,\left[{Q_0(x,y,t;k)}\right]\,\exp\,\left[{Q_0^*(x,y,t;k)}\right].
\end{gather}
Therefore, the $\tau$-funcion (\ref{Estevez:equation20}) is the  positive defined expression
\begin{gather}
\nonumber\tau_{1,2}=\left(\frac{X_1\left(X_1^2-3Y_1^2\right)}{3}\right)^2+\left(X_1^2Y_1-\frac{Y_1^3}{3}+8a_0t\right)^2+\\ +\quad \left(\frac{X_1^2+Y_1^2}{2a_0}\right)^2+\left(\frac{Y_1}{2a_0^2}\right)^2+\frac{1}{16a_0^2}.\label{Estevez:equation40}\end{gather}
The profile of this solution  is shown in Figure 3.
\begin{figure}
\begin{center}
\epsfig{file=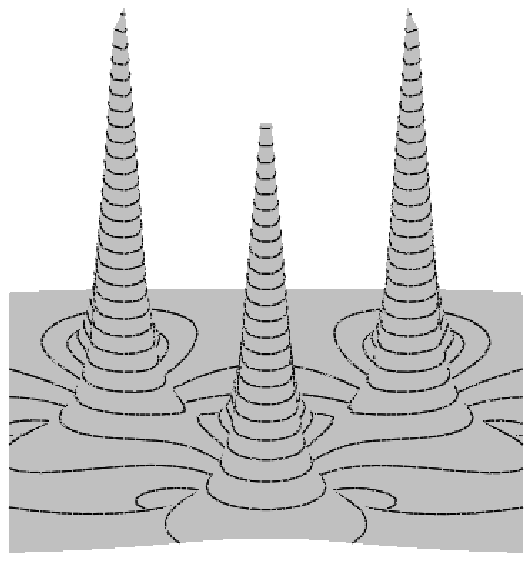,height=8cm}
\end{center}
\caption{Lump of the $1,1$ type}
\end{figure}
The asymptotic behavior of this solution can be obtained by considering the  transformation
\begin{gather}
\nonumber X_1=\hat X_1+c_1t^{\frac{1}{3}}\\
 Y_1=\hat Y_1+c_2t^{\frac{1}{3}}.
\label{Estevez:equation41}
\end{gather}
There are three possible solutions for $c_i$. For all of them $\tau_{1,2}$ is
\begin{equation}\tau{1,2}\rightarrow \hat X_1^2+\hat Y_1^2+\frac{1}{4a_0^2}\end{equation}

$\bullet$ $c_1=0,\, c_2=2(3a_0)^{\frac{1}{3}}$. This corresponds to a lump moving along the line
\begin{gather} X_1=\hat X_1\nonumber \\ Y_1=\hat Y_1+2(3a_0t)^{\frac{1}{3}}.
\end{gather}

$\bullet$ $c_1=-\sqrt 3\,(-3a_0)^{\frac{1}{3}},\, c_2=(-3a_0)^{\frac{1}{3}}$. This corresponds to a lump moving along the line
\begin{gather} X_1=\hat X_1-\sqrt 3\,(-3a_0t)^{\frac{1}{3}}\nonumber \\ Y_1=\hat Y_1+(-3a_0t)^{\frac{1}{3}}.
\end{gather}

$\bullet$ $c_1=\sqrt 3\,(-3a_0)^{\frac{1}{3}},\, c_2=2(3a_0)^{\frac{1}{3}}$. This corresponds to a lump moving along the line
\begin{gather} X_1=\hat X_1+\sqrt 3\,(-3a_0t)^{\frac{1}{3}}\nonumber \\ Y_1=\hat Y_1+(-3a_0t)^{\frac{1}{3}}.
\end{gather}

\section{Conclusions}
The Singular Manifold Method allows us to derive an iterative method to construct lump solutions characterized by two integers whose different combinations yield a rich possibilities of nontrivial self-interactions between the components of the solution.

+++++++++++++++

\subsection*{Acknowledgements}

This research has been supported in part by the DGICYT under project  FIS2009-07880.

\end{document}